\documentclass[sigconf]{acmart}
\usepackage{amsmath}
\usepackage[subrefformat=parens]{subcaption}
\usepackage{algorithmic}
\usepackage{algorithm}
\usepackage{booktabs}
\usepackage{multirow}

\def\model{\textsc{FluxCube}}

\makeatletter
\newcounter{parentsubcaption}
\newenvironment{subsubcaption}
 {\refstepcounter{sub\@captype}%
  \protected@edef\theparentsubcaption{\@nameuse{thesub\@captype}}%
  \setcounter{parentsubcaption}{\value{sub\@captype}}%
  \setcounter{sub\@captype}{0}%
  \@namedef{thesub\@captype}{\theparentsubcaption--\arabic{sub\@captype}}%
  \ignorespaces
}{%
  \setcounter{sub\@captype}{\value{parentsubcaption}}%
  \ignorespacesafterend
}
\makeatother

\usepackage{etoolbox}
\makeatletter
\patchcmd{\maketitle}{\@copyrightpermission}{
   \begin{minipage}{0.3\columnwidth}
     \href{https://creativecommons.org/licenses/by/4.0/}{\includegraphics[width=0.90\textwidth]{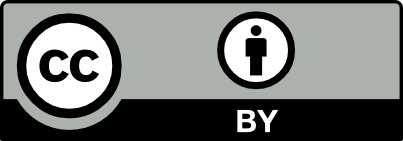}}
   \end{minipage}\hfill
   \begin{minipage}{0.7\columnwidth}
     \href{https://creativecommons.org/licenses/by/4.0/}{This work is licensed under a Creative Commons Attribution International 4.0 License.}
   \end{minipage}

   \vspace{5pt}
}{}{}

\makeatother

\AtBeginDocument{%
  \providecommand\BibTeX{{%
    \normalfont B\kern-0.5em{\scshape i\kern-0.25em b}\kern-0.8em\TeX}}}

\copyrightyear{2022}
\acmYear{2022}
\setcopyright{rightsretained}

\acmConference[CIKM '22]{Proceedings of the 31st ACM International Conference on Information and Knowledge Management}{October 17--21, 2022}{Atlanta, GA, USA}
\acmBooktitle{Proceedings of the 31st ACM Int'l Conference on Information and Knowledge Management (CIKM '22), Oct. 17--21, 2022, Atlanta, GA, USA}
\acmDOI{10.1145/3511808.3557396}
\acmISBN{978-1-4503-9236-5/22/10}

\begin{document}

\title{Mining Reaction and Diffusion Dynamics in Social Activities}

\author{Taichi Murayama}
\affiliation{%
  \institution{SANKEN, Osaka University}
  \country{Japan}}
\email{taichi@sanken.osaka-u.ac.jp}

\author{Yasuko Matsubara}
\affiliation{%
  \institution{SANKEN, Osaka University}
  \country{Japan}}
\email{yasuko@sanken.osaka-u.ac.jp}

\author{Yasushi Sakurai}
\affiliation{%
  \institution{SANKEN, Osaka University}
  \country{Japan}}
\email{yasushi@sanken.osaka-u.ac.jp}

\renewcommand{\shortauthors}{Taichi Murayama, Yasuko Matsubara, \& Yasushi Sakurai}


\begin{abstract}
Large quantifies of online user activity data, such as weekly web search volumes, which co-evolve with the mutual influence of several queries and locations, serve as an important social sensor.
It is an important task to accurately forecast the future activity by discovering latent interactions from such data, i.e., the ecosystems between each query and the flow of influences between each area.
However, this is a difficult problem in terms of data quantity and complex patterns covering the dynamics.
To tackle the problem, we propose \model, which is an effective mining method that forecasts large collections of co-evolving online user activity and provides good interpretability.
Our model is the expansion of a combination of two mathematical models: a reaction-diffusion system provides a framework for modeling the flow of influences between local area groups and an ecological system models the latent interactions between each query.
Also, by leveraging the concept of physics-informed neural networks, \model\ achieves high interpretability obtained from the parameters and high forecasting performance, together.
Extensive experiments on real datasets showed that \model\ outperforms comparable models in terms of the forecasting accuracy, and each component in \model\ contributes to the enhanced performance.
We then show some case studies that \model\ can extract useful latent interactions between queries and area groups.

\end{abstract}

\ccsdesc[500]{Information systems~Data mining}
\ccsdesc[500]{Mathematics of computing~Time series analysis}

\keywords{time series, data mininig, web data, reaction-diffusion systems, neural networks}

\maketitle

\section{Introduction}
The increasing volume of online user activity provides vital new opportunities to measure and understand the collective behavior of social and economic evolutions such as influenza prediction~\cite{ginsberg2009detecting}, the impact of individual performance~\cite{garimella2019hot}, user activity modeling~\cite{matsubara2012fast}, and other problems~\cite{leskovec2009meme,proskurnia2017predicting,zhao2013anatomy}.
A record of online user activity can play the role of a social sensor and offers important insights into people's decision-making.
In particular, accurately forecasting the volume of online user activity and unraveling hidden patterns and interactions within them have many benefits.
For example, marketers want to know the future popular volume of products and the relationship between multiple products and locations in online user attention to enable appropriate advertisement and new product development.
They can avoid wasting human and material resources by accurately modeling future behavior.

The modeling of online user activity remains a challenging task due to the increasing data volume, which has multiple domains and a network of time series with mutual influences.
For example, when we consider online user behavior analysis using web search activities as observations, which could be in the form of 3rd-order tensor (timestamp, location, keyword), it is important to deal with the following problems: 
(a) \textbf{\textit{Capturing latent interactions and diffusion between observable data}}: 
Many time series for individual locations and keywords are not independent.
For example, the search volume per keyword in the same category may compete for user resources~\cite{matsubara2015web}.
Also, the popularity of a keyword in one location diffuses and affects the search volume in another location~\cite{okawa2021dynamic}.
In other words, we consider changes in the popularity of a keyword have the properties of flux phenomena, that can be expressed as a flow of influence.
(b) \textbf{\textit{Capturing latent temporal patterns behind observable data}}: 
Many time series contain several patterns, such as trends and seasonality.
Such patterns behind time series reveal the relationship between people's activities, such as Black Friday.
However, it is extremely difficult to design an appropriate model for such patterns by hand without knowing their real characteristics in advance.
The modeling method should therefore be fully automatic as regards estimating the hidden pattern for understanding data structures and saving human resources.
(c) \textbf{\textit{Accurate forecast}}: Accurate time series forecasts give marketers useful insights into the future trends of their keywords and avoid wasting human and material resources. 
The key challenge is to achieve high forecasting accuracy while addressing the two aforementioned issues related to the interpretability of observable data.

\begin{figure*}[t]
  \begin{subfigure}[b]{0.28\linewidth}
    \centering
    \includegraphics[width=\textwidth]{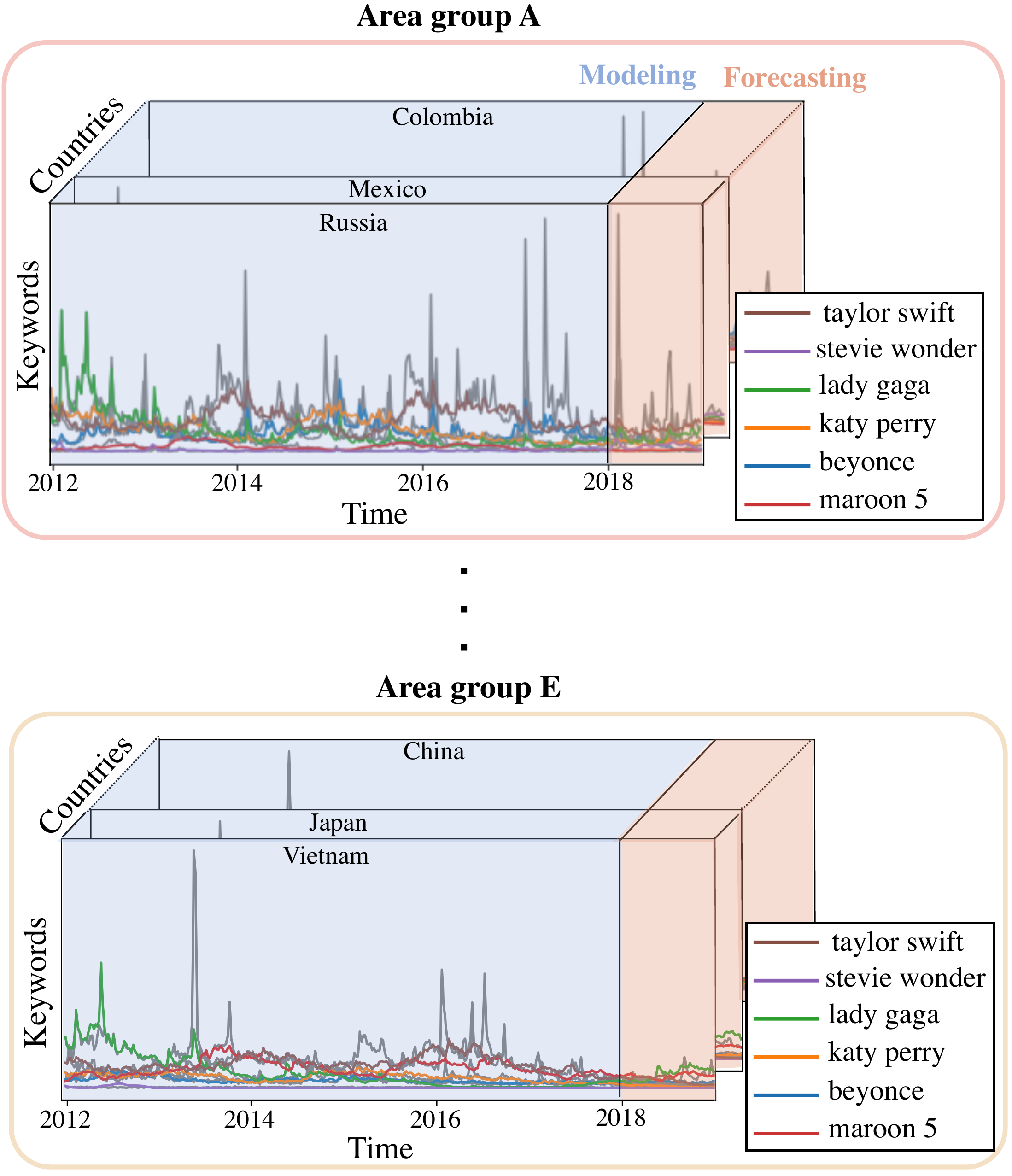}
    \subcaption{Modeling and forecasting tensor data of each area group}
  \end{subfigure}
  \hfill
  \begin{subfigure}[b]{0.20\linewidth}
    \centering
    \includegraphics[width=\textwidth]{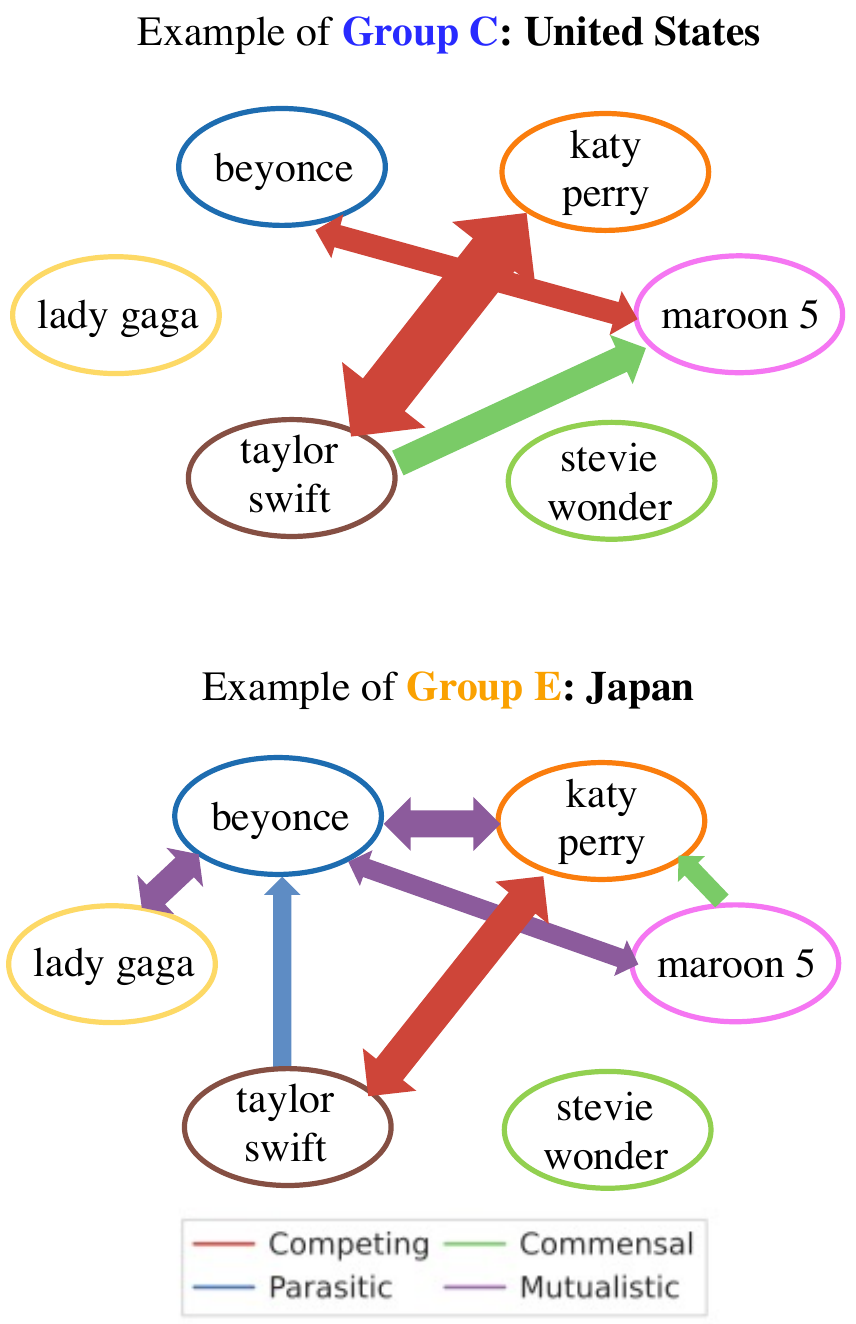}
    \subcaption{Latent interactions}
  \end{subfigure}
  \hfill
  \begin{subfigure}[b]{0.50\linewidth}
    \centering
    \includegraphics[width=\textwidth]{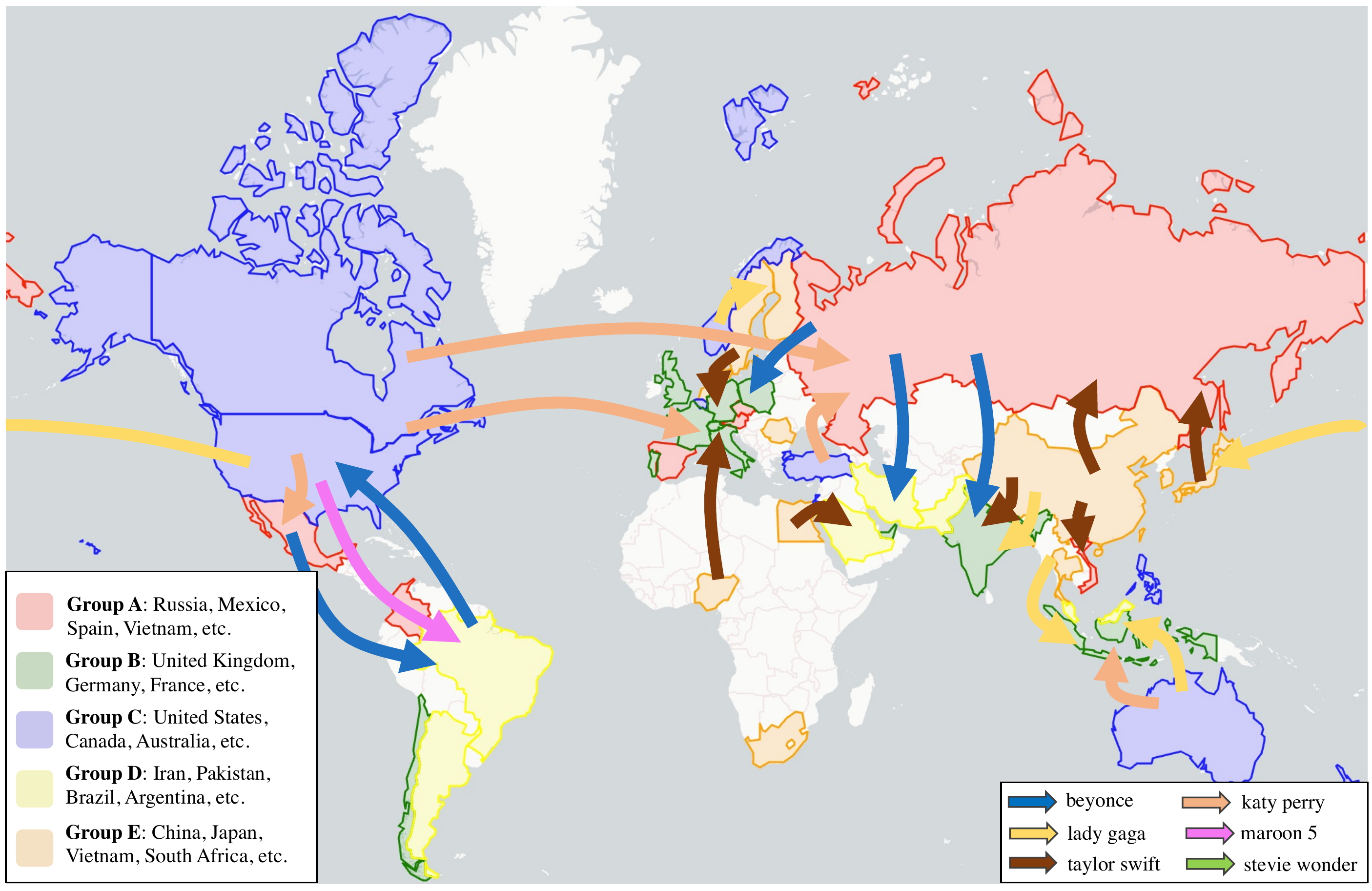}
    \subcaption{Diffusion process of each area group}
  \end{subfigure}
  \caption{Modeling power of \model\ for online user activity data related to six artists (World\#1): (a) Given the original data (gray lines), \model\ captures the dynamics of social activities (blue). Futhermore, \model\ provides a multiple steps ahead forecast (red). (b) Our method detects the hidden interactions between keywords in any country. The arrows indicate the direction and the type of influence. (c) It also automatically discovers similar area groups based on their keyword interactions and then finds the flow of influence of each keyword between area groups. The figure visualizes the flow of influence in 2015.
  }
  \label{overview_figure}
\end{figure*}

This paper focuses on an important time series modeling and forecasting task, and we design \model, which is an effective mining method that forecasts large collections of co-evolving online user activity and provides good interpretability.
This model solves (a) and (b) in the above problems with the combination of mathematical models: a reaction-diffusion system~\cite{holmes1994partial} representing the change in space and time of chemical substances and the Lotka-Volterra population model~\cite{marsden2002interdisciplinary,haefner2005modeling} describing the population dynamics of prey and predator.
Also, it solves (c) with the idea of physics-informed neural networks~\cite{raissi2019physics} to conduct supervised learning tasks while respecting any given laws of physics described by general nonlinear partial differential equations.
Intuitively, the problem we wish to solve is as follows.

\textsc{Informal Problem 1.} 
\textit{\textbf{Given} a tensor $\mathcal{X}^{c}$ up to the time point $t_{c}$, which consists of elements at $L$ locations for $K$ keywords, i.e., $\mathcal{X}^{c} = \{x_{tij}\}^{t_{c},L,K}_{t,i,j = 1},$}
\begin{itemize}
    \item \textit{\textbf{find} trends and seasonal patterns}
    \item \textit{\textbf{find} area groups of similar dynamics}
    \item \textit{\textbf{find} latent interections between keywords}
    \item \textit{\textbf{find} the flow of influence between area groups}
    \item \textit{\textbf{forecast} $l_{f}$-step ahead future values, i.e., $\mathcal{X}^{f} = \{x_{tij}\}$ where ($t = t_{c} + l_{f}; i = 1,...,L;j = 1,...,K$)}
\end{itemize}

\subsection{Preview of our results}
\begin{figure}[t]
    \centering
    \includegraphics[width=\linewidth]{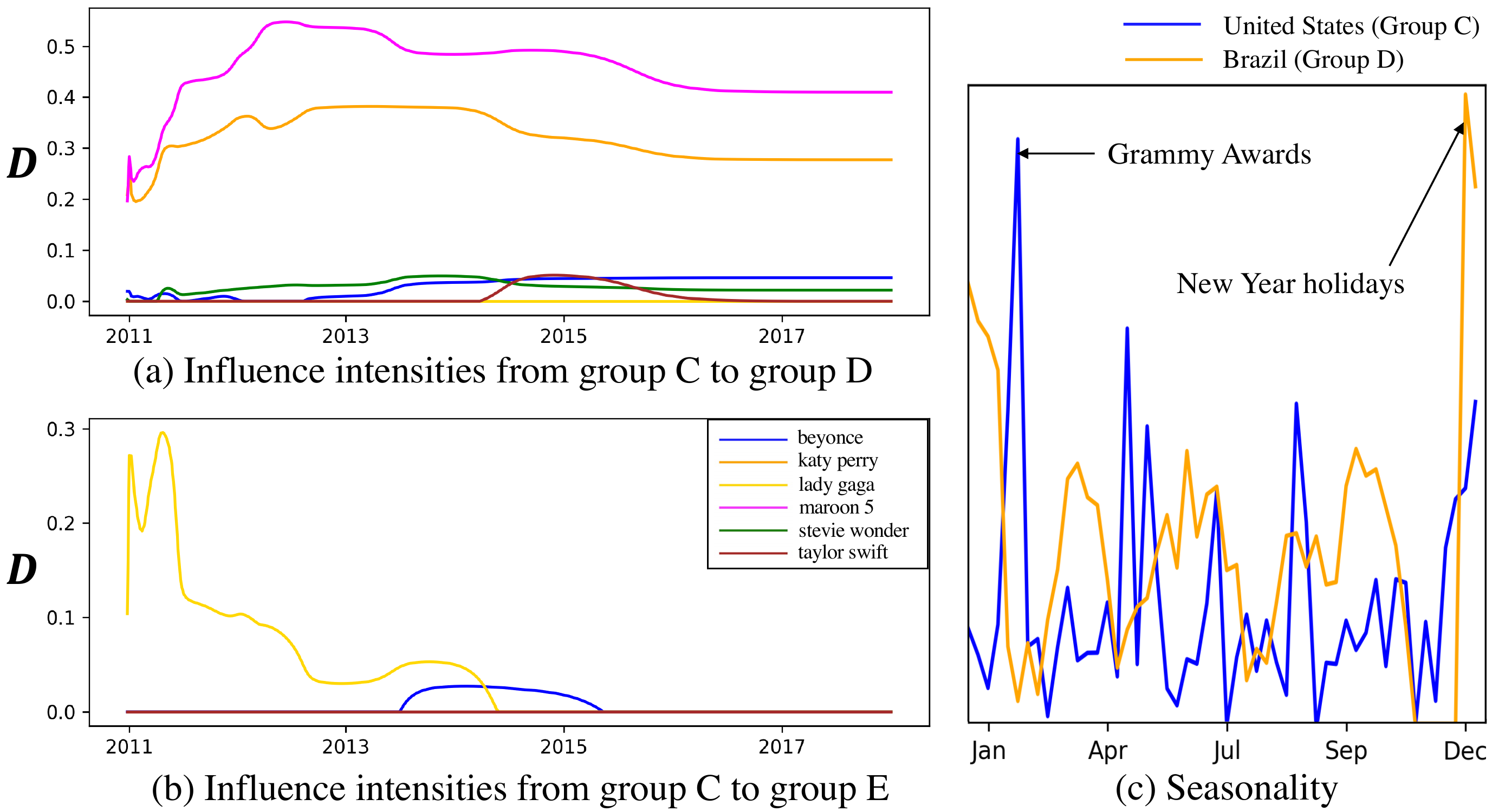}
    \caption{Latent pattern mining of \model\ for search volumes related to six artists in the US: (a) and (b) show that it automatically detects time-varying influence intensities related to each artist from group C. (c) shows that it automatically detects seasonality patterns about keyword ``beyonce'' in the US and Brazil.}
    \label{overview_figure2}
\end{figure}

Figures~\ref{overview_figure} and \ref{overview_figure2} show the results obtained with \model\ for modeling online user activity data related to six artists (World\#1). 
Specifically, our method captures the following properties:
\begin{itemize}
    \item \textbf{Long-term trends}: Figure~\ref{overview_figure} (a) shows tensor data that consists of weekly online search volumes in relation to artists in fifty countries.
    \model\ automatically captures the trends and the hidden interactions between keywords and area groups in the modeling term (blue part) and realizes the accurate long-term forecasting over two years (red part).
    
    \item \textbf{Latent interaction between keywords}: 
    Figure~\ref{overview_figure} (b) shows the latent interaction between each keyword. 
    \model\ uncovers four types of interaction (competing, commensal, parasitic, mutualistic) between the keywords, where the arrow colors and widths show the interaction type and its connection intensity, from the tensor data in each location.
    For example, while the relationship between maroon 5 and beyonce is competing in the United States, it is mutualistic in Japan.
    Also, the competing relationship between taylor swift and katy perry is common in the two countries.
    
    \item \textbf{Diffusion process of each area group}: 
    \model\ finds the location clustering and the flow of influence between each area group, as shown in Figure~\ref{overview_figure} (c). 
    This figure shows that the colors of the countries on the map indicate area groups, and the colors and directions of the arrows indicate the keyword type and the influence flow, respectively.
    \model\ discovers the most suitable clustering result: Europe in group B, North America in group C, and East Asia in group D. 
    The influences of each keyword on each group, which are represented by arrows, have time-varying intensities, which \model\ also captures, as shown in Figures~\ref{overview_figure2} (a) and (b).
    There are constant influence flows of maroon 5 and katy perry from group C to D, while there is an influence flow of taylor swift in the early 2010s from group C to E, which has disappeared.
    Our model provides the flow of influence in an intuitive form.
    
    \item \textbf{Sesonality}: 
    Figure~\ref{overview_figure2} (c) shows the seasonalities extracted from the search volume data. 
    Two kinds of annual patterns resulting from ``Grammy Awards'' and ``New Year holidays'' are found from search volumes in the United States and Brazil.
    It is important to discover such periodic patterns if we are to accurately forecast and model the dynamics.
\end{itemize}

\section{Related Work}
In this section, we briefly describe investigations related to our research.
The details of related work are separated into three categories: time series forecasting and modeling, web data modeling, and a physics-informed neural network.

\textbf{Time series forecasting and modeling}: 
There is a lot of interest in time series forecasting and modeling.
Conventional approaches to time series forecasting and modeling are based on statistical models such as auto regression (AR), state space models (SSM), Kalman filters (KF), and their extensions~\cite{durbin2012time,dabrowski2018state,li2010parsimonious}.
Some models combining these classical methods with dimension reduction, such as TRMF~\cite{yu2016temporal} and SMF~\cite{hooi2019smf} have shown useful results in the field of data mining.
However, these methods are limited as regards the interpretability of observation data because of the dimension reduction.
In recent years, many models have been based on neural networks thanks to their rapid development.
These networs include convolutional neural networks (CNNs)~\cite{wavenet,cnn1,cnn2} and recurrent neural networks (RNNs)~\cite{salinas2020deepar,rnn_state,rnn_ss2}, which capture the temporal variation.
In particular, time series forecasting models using Transformer, developed in the field of a natural language preocessing~\cite{transformer_original}, achieve high prediction performance~\cite{attention_apply1,attention_apply2,attention_apply3}.
For example, Informer~\cite{zhou2021informer}, one of the Transformer-based models, achieves a successful outcome in long-ahead forecasting.
Deep learning-based models guarantee high forecasting performance; however, they are still problematic in terms of their interpretability.
We achieve high interpretability as well as a high predictive performance by our mathematical model fused with deep learning.

\textbf{Web data modeling}: 
Online user activity data, such as web search query data and social media posts, are expected to capture human movements as social sensors~\cite{chang2021supporting,aramaki2011twitter,ribeiro2021sudden} and are utilized in real-world applications such as influenza forecasting~\cite{ginsberg2009detecting}, talent flow forecasting~\cite{zhang2019large}, and item popularity prediction~\cite{mishra2016feature}.
This potential for real-world application has led to a huge interest in uncovering and modeling the movement of web data.
The author of \cite{beutel2012interacting,prakash2012winner} investigated the competing dynamics of two keywords on the web and found factors impacting final states based on the infection model.
Various other modeling studies on the web have been conducted as follows: competing dynamics of membership-based websites~\cite{ribeiro2014modeling}, modeling the content and viewer dwell time on social media platforms~\cite{lamba2019modeling}, modeling the online activities into return and exploration across online communities~\cite{hu2019return,li2019fates}, extracting the relationships between search queries and external events~\cite{karmaker2017modeling,karmaker2018jim}, and modeling online user behavior patterns in the knowledge community~\cite{mavroforakis2017modeling}.
CubeCast~\cite{kawabata2020non} and EcoWeb~\cite{matsubara2015web} are focused on modeling the dynamics of web search queries.
CubeCast is an online algorithm designed to capture useful time series segmentation and seasonality patterns.
EcoWeb is a modeling method for capturing keyword interactions from the dynamics of web search queries based on biological mathematics.
These studies revealed the interaction of search queries in a single location, but did not explore how keywords interact between locations.
Our proposed model discovers the flow of influence between locations from the time series of each query.

\textbf{Physics-informed neural network}:
Recently, neural networks have found applications in the context of numerical solutions of partial differential equations and dynamical systems because a neural network is a universal approximator~\cite{raissi2019physics}.
These models, which are called physics-informed neural networks, can solve forward problems~\cite{raissi2018hidden}, where the approximate solutions of governing equations are obtained, as well as inverse problems, where parameters involved in the governing equation are obtained from the observation data~\cite{fang2019physics}.
We try to utilize the idea to represent some of the parameters in our partial differential equations flexibly with a neural network.


\begin{table}
\footnotesize
   \caption{Symbols and definitions} 
  \label{symbol}
  \centering
  
  \begin{tabular}{l|l} \toprule
  Symbol & Definition\\ \midrule
  $t_{c}, t_{f}$ & Modeling and forecasting time point\\
  $\mathcal{X}^{c}, \mathcal{X}^{f}$ & Modeling and forecasting tensor data\\
  $l_{c}, l_{f}$ & Time interval for modeling and forecasting\\
  $L, K$ & Number of locations and keywords\\
  $d_{l}$ & Number of area groups\\ \midrule
  $a_{j}$ & Intrinsic growth rate of keyword $j$\\
  $b_{j}$ & Carrying capacity of keyword $j$\\
  \multirow{2}{*}{$c_{jj'}$} & Intra/inter-keywords interaction strength from $j'$-th keyword\\
   & to $j$-th keyword\\
  $\boldsymbol{D}^{t}_{mnk}$ & Influence flow of the keyword $k$ from location $n$ to $m$\\
  $\boldsymbol{S}$ & Seasonality latent pattern matrix\\
  $p$ & Period of seasonality\\
  $\alpha$, $\beta$ & Weights of regularization terms for loss function\\
  $\boldsymbol{o}^{i}$ & Compression representations of item interactions in location $i$\\
  $g^{1},...g^{d_{l}}$ & Area groups \\
  $\Theta'_{d_{l}}$ & Parameter sets related to $d_{l}$\\
  \bottomrule
  \end{tabular}
\end{table}

\section{Overview}
This section provides an overview of our proposed model, namely \model, for mining and forecasting co-evolving online user activity data.
We first introduce related notations and definitions and then describe the characteristics of \model.

\subsection{Problem definition}
Table~\ref{symbol} lists the main symbols that we use throughout this paper.
We consider online user activity data to be a 3rd-order tensor, which is denoted by $\mathcal{X} \in \mathbb{R}^{T \times L \times K}$, where $T$ is the number of time points, and $L$ and $K$ are the numbers of locations and keywords, respectively.
The element $x_{tij}$ in $\mathcal{X}$ corresponds to the search volume at time $t$ in the $i$-th location of the $j$-th keyword.
Our overall aim is to realize the long-term forecasting of a tensor $\mathcal{X}$ while extracting the hidden patterns and latent interactions.
We define $\mathcal{X}^{c} =  \{x_{tij}\}^{t_{c},L,K}_{t,i,j = 1}$ as observable tensor data, whose length is denoted by $l_{c}$.
Similarly, let $\mathcal{X}^{f} =  \{x_{tij}\}^{t_{f},L,K}_{t,i,j =t_{c},1,1}$ denotes a partial tensor from $t_{c}$ to $t_{f}$ for forecasting.
Here, we define $l_{c}$ and $l_{f}$ as a time interval for modeling and forecasting.
Finally, we formally define our problem as follows.

\textsc{Problem 1 ($l_{f}$-step ahead forecasting).} 
\textit{\textbf{Given} a tensor $\mathcal{X}^{c} = \{x_{tij}\}^{t_{c},L,K}_{t,i,j = 1},$ up to the time point $t_{c}$; 
\textbf{Forecast}: $l_{f}$-step ahead future values  $\hat{\mathcal{X}}^{f} =  \{\hat{x}_{tij}\}^{t_{f},L,K}_{t,i,j =t_{c},1,1}$, where $t_{f} = t_{c} + l_{f}$.}

\subsection{Reaction-diffusion system}
Our model for capturing the latent interactions between keywords and area groups is inspired by a reaction-diffusion system.
A reaction-diffusion system is a mathematical model corresponding to physical phenomena such as the change in space and time of chemical substances, represented as partial differential equations.
The system can also describe the dynamic processes of non-chemical areas such as biology, geology, and physics~\cite{fitzhugh1961impulses,ganguly2017reaction,pearson1993complex,cosner2008reaction}.
The general form of a reaction-diffusion system can be described with the following equations:
\begin{equation}
    \frac{\partial \boldsymbol{u}}{\partial t}=  f(\boldsymbol{u}, t) + \mathcal{D}\Delta\boldsymbol{u} \label{reaction-diffusion}
\end{equation}
where $\boldsymbol{u}$ represents the concentration values in a chemical substance, $t$ represents the present time, and $\mathcal{D}$ is the diffusion coefficient.
The first term on the right hand side, $f(\boldsymbol{u}, t)$, represents the reaction term, which accounts for all local reactions, and the second term, $\mathcal{D}\Delta\boldsymbol{u}$, represents the diffusion term, which accounts for how the substance spreads to other locations.

Our model is an extended reaction-diffusion system for modeling and forecasting online user activity data.
For example, the related search volumes in the data interact with each other and compete for user resources.
The interaction between keywords in a location can be represented as a reaction term in the reaction-diffusion system.
Also, a vogue for any keyword in a location can affect that in another location.
Such an influence flow of any keyword between locations can be represented as a diffusion term.
Thus, we assume that the interactions between area groups and keywords can be represented by a diffusion-reaction system.

\subsection{Capabilities}
To capture all the components outlined in the introduction, we present our model, namely \model\, which can model the latent patterns and interactions underlying online user activity 3rd-order tensor data.
So, how can we build our model so that it models the user activity data while capturing the latent patterns and interactions and achieving accurate forecasting?
Specifically, our model should have the following three capabilities: 
\begin{itemize}
    \item \textbf{Latent interaction and diffusion}: User activity data evolves naturally over time depending on many latent factors, such as user preferences and customs for online web-search activities.
    We need to capture the latent interactions of keywords and the influence flow of locations from the real-time series data.
    To explicitly capture such interactions, we propose using a reaction-diffusion system~\cite{holmes1994partial} including the Lotka-Volterra population model~\cite{marsden2002interdisciplinary}, which is expressed in differential equations.
    
    \item \textbf{Seasonality}: We should also note that online user activity has certain annual patterns. 
    As Figure~\ref{overview_figure2} (c) shows, there are annual patterns such as a huge spike in New Year holidays in online user activity data.
    Capturing seasonalities realizes our suitable modeling.
    
    \item \textbf{Time-varying components}: User activity data change significantly due to external events such as the influence of other locations.
    The variables in the differential equation cannot easily explain the relationships between locations that vary greatly with time and external events, for which we require a flexible modeling method.
    To achieve this, we leverage the idea of physics-informed neural networks~\cite{raissi2019physics}.
    Intuitively, our model combines a neural network with the characteristics of universal approximation~\cite{liang2017deep} and differential equations with high interpretability.
\end{itemize}

\begin{figure}
    \centering
    \includegraphics[width=\linewidth]{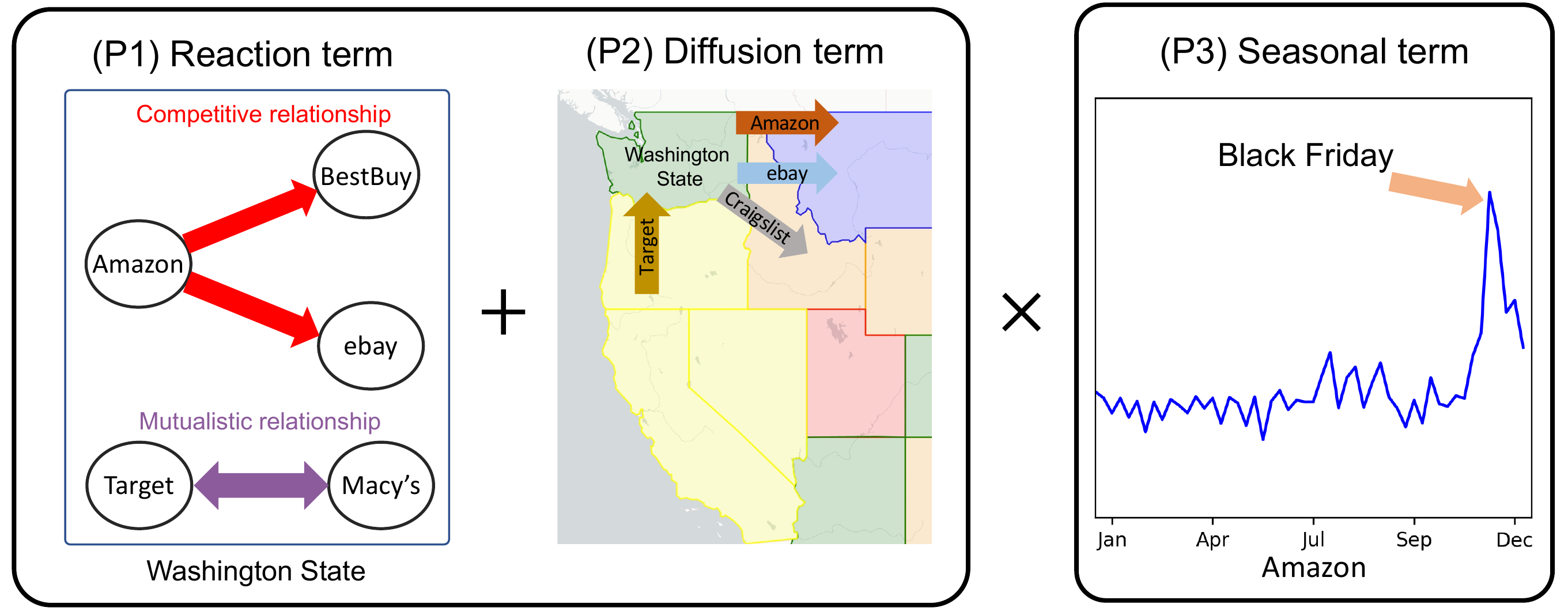}
    \caption{Conceptual image of our model targeting ecommerce data (US\#1) in Washington state. Our model consists of three parts: (P1) Reaction term represents the interaction between keywords, (P2) Diffusion term represents the influence flow of each keyword between locations, and (P3) Seasonal term represents the seasonality of each keyword.}
    \label{model_overview}
\end{figure}

\section{\model}
Here, we describe the formula of \model.
We represent the base form of our model applied to online user activity data, mainly web search volumes, which is inspired by the reaction-diffusion system, Eq. (\ref{reaction-diffusion}), in the following:
\begin{align}
    \frac{\partial x_{tij}}{\partial t} &=  f(x_{tij}|\mathcal{X}^{c}_{t,i,{:}}) + g(x_{tij}|\mathcal{X}^{c}_{t}, t) \label{base1} \\
    \hat{x}_{t+1ij} &= F(\frac{\partial x_{tij}}{\partial t} + x_{tij}) \label{base2}
\end{align}
The first term in Eq. (\ref{base1}) serves as the reaction term for extracting the interaction between keywords in location $i$ (\textbf{P1}).
The second term in Eq. (\ref{base1}) serves as the diffusion term for extracting the influence flow of keyword $j$ between locations (\textbf{P2}).
Finally, we set Eq. (\ref{base2}) to capture seasonality (\textbf{P3}).
A conceptual drawing of our modeling method is shown in Figure~\ref{model_overview}. 

Next, we introduce each component of the above form of our model in steps.

\subsection{Reaction term (P1)}
We first introduce the reaction term in detail.
The term aims to model the interaction between keywords within a location.

Observation of the search volume of each keyword on the web shows that the keywords compete for user attention.
No keyword can survive on the web if no one pays attention to that topic.
Besides, the number of instances of user attention on the web is finite.
This relationship between keywords and user attention on the web is very similar to the relationship between species and food resources in the jungle, e.g., squirrel monkeys and spider monkeys compete for fruit and no species can survive without resources.
We employ the Lotka-Volterra population model~\cite{marsden2002interdisciplinary,haefner2005modeling}, which represents the growth of the species considering the interaction between them in biological mathematics, as the reaction term.

The reaction term in Eq. (\ref{base1}) for keyword $j$ is represented below:
\begin{equation}
    f(x_{tij}|\mathcal{X}^{c}_{t,i,{:}}) = a_{j}x_{tij}\left(1 - \frac{\Sigma_{j'}{c_{jj'}x_{tij'}}}{b_{j}}\right),  \label{lotka}
\end{equation}
where, $a_{j} > 0$, $b_j > 0$ and $c_{jj} = 1$.
The size of each parameter depends on the number of keywords $K$: $\boldsymbol{a} \in \mathbb{R}^{K}$, $\boldsymbol{b} \in \mathbb{R}^{K}$, and $\boldsymbol{C} \in \mathbb{R}^{K \times K}$.
Each parameter is interpreted as:
\begin{itemize}
    \item $a_{j}$: intrinsic growth rate of keyword $j$;
    \item $b_{j}$: carrying capacity of keyword $j$;
    \item $c_{jj'}$: intra/inter-keyword interaction strength from the $j'$-th keyword to the $j$-th keyword, which is each value in $\boldsymbol{C}$
\end{itemize}
The signs of the off-diagonal elements of $\boldsymbol{C}$ provide an interpretation of the latent relationship between the two keywords:
\begin{itemize}
    \item $c_{jj'} > 0, c_{j'j} > 0$: a competitive relationship
    \item $c_{jj'} < 0, c_{j'j} > 0$: a parasitic relationship
    \item $c_{jj'} < 0, c_{j'j} = 0$: a commensal relationship
    \item $c_{jj'} < 0, c_{j'j} < 0$: a mutualistic relationship
\end{itemize}
We expect these characteristics to be available for modeling online user activity on the web and extracting the relationships between keywords.

\subsection{Diffusion term (P2)}
We then introduce the diffusion term in detail.
The term aims to model the influence flow of any keyword between locations.

Can partial differential equations and mathematical models adequately represent the relationships between locations in online user activity on the web?
This task is hard to achieve.
The reaction-diffusion system in chemistry can represent the change in space and time of chemical substances by a constant such as the diffusion coefficient, based on the observation that the spread of a chemical substance is constant without external influences.
However, on the web, interactions of any keyword between locations are not constant due to external factors.
To capture the time change of the interaction, a dynamic linear model with time-varying coefficients could be utilized, but it is difficult to take account of complex phenomena and rapid changes.
Here, we utilize a neural network with universal approximation as part of our mathematical model to represent the changing interactions between locations over time.
This corresponds to a kind of physics-informed neural network~\cite{raissi2019physics}.
Concretely, we represent the diffusion term in Eq. (\ref{base1}) for keyword $j$ at location $i$ as below:
\begin{align}
    \boldsymbol{D}^{t} =& ReLU\left(RNN(1:t)\right) \label{react1}\\
    \boldsymbol{E}^{t}_{i} =& \boldsymbol{D}^{t}_{i} \odot \mathcal{X}_{t},  \qquad  (i = 1,...,d_{l}) \label{react2}\\
    g(x_{tij}|\mathcal{X}^{c}_{t}, t) =& \sum^{d_{l}}_{i'} \boldsymbol{E}^{t}_{i,i',j} \label{react3}
\end{align}
where $\boldsymbol{D}^{t} \in \mathbb{R}^{d_{l} \times d_{l} \times K}$ and $\boldsymbol{E}^{t} \in \mathbb{R}^{d_{l} \times d_{l} \times K}$, and $\boldsymbol{D}^{t}_{i,i,:} = 0$.
$\boldsymbol{D}^{t}_{i} \in \mathbb{R}^{d_{l} \times K}$ and $\boldsymbol{E}^{t}_{i} \in \mathbb{R}^{d_{l} \times K}$ represent the $i$-th vectors of $\boldsymbol{D}^{t}$ and $\boldsymbol{E}^{t}$, respectively.
We express the flow of attention of each user between locations using a recurrent neural network (RNN)~\cite{rumelhart1986learning}, which captures time-varying parameters using time $1:t$ as input in Eq. (\ref{react1}).
The RNN enables our model to be more expressive by allowing time-varying parameters along with temporal dependencies in the covariates. 
The output of RNN in Eq. (\ref{react1}) is transformed to a tensor form: $\boldsymbol{D}^{t}$ is a form of 3rd-order tensor and $\boldsymbol{D}^{t}_{mnk}$ indicates the degree to which location $n$ contributes to the popularity of the keyword $k$ in location $m$.
In other words, $\boldsymbol{D}^{t}_{mnk}$ represents the influence intensities of the keyword $k$ from $n$ to $m$, which is provided in an intuitive form for us.
We show some examples of $D^{t} (t = 1,...,t_{c})$ in Figure~\ref{base2} (a) and (b).
A total of the influences that keyword $j$ at location $i$ at time $t$ receives from other locations is expressed by multiplying the observed data $\mathcal{X}$ with the output in Eq. (\ref{react1}) ($\boldsymbol{D}$), as shown in Eq. (\ref{react2}) and (\ref{react3}).
By applying a neural network to a part of our mathematical model, we expect to achieve both flexible modeling and high explainability.

\subsection{Seasonality (P3)}
We finally consider seasonal/cyclic patterns in user activity data by extending our model.
Each keyword always has certain users' attention; however, the users change their behavior dynamically, according to various seasonal events, e.g., Christmas Day and Black Friday.
We need to detect hidden seasonal activities with our model.
More specifically, we model the user activity value at the next step considering the seasonality pattern from our reaction-diffusion system in Eq. (\ref{base2}), as below:
\begin{align}
    F\left(\frac{\partial x_{tij}}{\partial t} + x_{tij}\right) = \left(\boldsymbol{1} + [\boldsymbol{S}_{t \bmod p}]_{i,j}\right) \odot \left(\frac{\partial x_{tij}}{\partial t} + x_{tij}\right) \label{season}
\end{align}
where $\boldsymbol{S} \in \mathbb{R}^{p \times L \times K}$, $\boldsymbol{S} \geq 0$, $p$ is period of seasonality, and $\odot$ indicates the element-wise product.
$\boldsymbol{S}$ acts as a projection matrix for latent seasonal patterns, such as the seasonal term in Figure \ref{model_overview}.

\subsection{Loss function}
To train \model, we use the mean squared error (MSE) loss with regularization terms as below:
\begin{align}
    \|\mathcal{X}^{c} - \hat{\mathcal{X}}\|^{2} + \alpha\sum\|\boldsymbol{D}\|^{2} + \beta\sum\|\boldsymbol{S}\|^{2} \label{loss}
\end{align}
where $\alpha$ and $\beta$ are hyper-parameters for controling the weights of the regularization terms, and $\hat{\mathcal{X}}$ is our modeling value.
The first term is MSE loss that penalizes the difference between modeling and observation values, and the second and third terms are regularization terms that encourage the suppression of the adoption of large values for two parameters: the influence between locations ($\boldsymbol{D}$) and the latent seasonal patterns ($\boldsymbol{S}$).

\section{Optimization algorithm}
This section presents our optimization algorithm for learning \model\ while grouping locations with similar user activity characteristics. 

Thus far, we have proposed \model\ for modeling and forecasting online user activity data with multiple locations.
There are two challenges in fitting our model to data with many locations: (a) the learning can be inefficient and unstable because there are large candidate parameters that must be estimated and converged, and (b) the interactions between locations become more complex and results less interpretable.
Here, we propose an algorithm to learn \model\ where we cluster locations that have observed data with similar characteristics as the same area group.

\subsection{Automatic location clustering}
Our hypothesis for clustering online user activity data as regards location is simple.
It is that locations with similar interactions between items represented by the reaction term (P1) can be considered as the same area group.
The clustering method refers to \cite{darwish2020unsupervised}, according to the number of area group $d_{l}$ we cluster the locations, the number of which is $L$, using Uniform Manifold Approximation and Projection (UMAP)~\cite{mcinnes2018umap} and K-means:
The equations are as below:
\begin{align}
    \boldsymbol{o}^{i} &= UMAP(\boldsymbol{a}^{i}, \boldsymbol{b}^{i}, \boldsymbol{C}^{i}) \qquad  (i = 1,...,L) \label{umap_eq} \\ 
    g^{1},...,g^{d_{l}} &= \text{K-means}(\boldsymbol{o}, d_{l}) \label{kmeans_eq}
\end{align}
where $\boldsymbol{o}^{i} \in \mathbb{R}^{2}$, $\boldsymbol{o} \in \mathbb{R}^{L \times 2}$ is the compression representation of the keyword interactions and $L$ is the number of locations.
We perform clustering using three parameters ($\boldsymbol{a}^{i}, \boldsymbol{b}^{i}, \boldsymbol{C}^{i}$) that represent the interaction of keywords at location $i$ in Eq. (\ref{lotka}) in the reaction term. 
These three parameters are compressed to the two-dimensional vector $\boldsymbol{o}^{i}$ by UMAP.
We then cluster the compression representation $\boldsymbol{o}$ by K-means into $d_{l}$ area groups ($g^{1},...,g^{d_{l}}$).

\subsection{Finding the appropriate number of area groups}
We describe how to find the appropriate number of area groups $d_{l}$.
The situation where our model with a small number of parameters including $d_{l}$ can adequately explain the observed data is the best.
Here, we apply the minimum description length (MDL) principle~\cite{chakrabarti2004fully,tang2003mining,rissanen1978modeling} to find the appropriate $d_{l}$.
The MDL principle enables us to determine the nature of good summarization by minimizing the sum of the data encoding cost and the model description cost.
The cost is described as below:
\begin{align}
    <\mathcal{X}^{c} | \Theta'> + < \Theta'_{d_{l}}> + \ C \label{encoding}
\end{align}
where $<\mathcal{X}^{c} | \Theta'>$ represnts the cost of describing the data $\mathcal{X}^{c}$ given the model parameters $\Theta'$, $< \Theta'_{d_{l}}>$ shows the cost of describing $\Theta'_{d_{l}}$, which is a parameter set that varies with $d_{l}$, and $C$ is a constant value, which is not affected by $d_{l}$.
In short, it follows the assumption that the more we compress data, the more we can learn about the underlying pattern.
We begin by defining the two components of the total cost more concretely.

\noindent \textbf{Data encoding cost}: 
We can encode the observation data $\mathcal{X}^{c}$ using $\Theta'$ based on Huffman coding~\cite{rissanen1978modeling}.
The coding scheme assigns a number of bits to each value in $\mathcal{X}^{c}$, which is the negative log-likelihood under a Gaussian distribution with mean $\mu$ and variance $\sigma^{2}$, i.e.,
\begin{align}
    <\mathcal{X}^{c} | \Theta'> = \sum_{t,i,j = 1}^{t_{c}, L, K} -\log_2 {p_{\mu,\sigma}(x^{c}_{t,i,j} - \hat{x}^{c}_{t,i,j})}
\end{align}
where, $\hat{x}^{c}_{t,i,j} \in \hat{\mathcal{X}}^{c}$ is the modeling value of $x^{c}_{t,i,j} \in \mathcal{X}^{c}$ by our proposed model.

\noindent \textbf{Model description cost}:
The model cost is the number of bits needed to describe the model.
If we use a more powerful model architecture, the total cost becomes higher.
We focus only on parameters related to $d_{l}$, i.e., $\boldsymbol{D}$ in the diffusion term, because of our objective of selecting the appropriate $d_{l}$.
Parameters unrelated to $d_{l}$ can be treated as constant value $C$, which we can ignore in the cost calculation.
Our model cost is represented as below:
\begin{align}
    < \Theta'_{d_{l}}> &=  < d_{l} > \ + \ < D > \\
    < d_{l} > &= \log^{*}(d_{l})\\
    < D > &= |\boldsymbol{D}| \cdot (\log(t_{c})+2\cdot\log(d_{l})+\log({K}) + c_{F}) + \log^{*}(|\boldsymbol{D}|)
\end{align}
Here, $\log^{*}$ is the universal code length for integers, $|\cdot|$ describes the number of non-zero elements and $c_{F}$ denotes the floating point cost (32 bits).

\begin{figure}[!t]
\begin{algorithm}[H]
    \caption{Optimization} 
    \label{alg1} 
    \begin{algorithmic}[1]
    \REQUIRE Input tensor $\mathcal{X}^{c}$
    \ENSURE Learned \model\ with interpretable parameters
    \STATE /* \textbf{Initialize} 
    \STATE $d_{l} \leftarrow 1$
    \STATE $g^{d_{l}} \leftarrow \{1,...,L\}$
    \STATE $\text{MinCost} \leftarrow \text{inf}$
    \WHILE{\text{improving the cost}}
    \IF{$d_{l} > 1$}
    \STATE /* Automatic location clustering (Section 5.1) 
    \STATE $g^{1},...,g^{d_{l}} \leftarrow Clustering(d_{l}, \boldsymbol{a}_{d_{l}-1}, \boldsymbol{b}_{d_{l}-1}, \boldsymbol{C}_{d_{l}-1})$
    \ENDIF
    \STATE /* \model\ training (Section 4)
    \STATE $\model_{d_{l}} \leftarrow Training(\mathcal{X}^{c}, g^{1},...,g^{d_{l}})$
    \STATE /* Calculate the encoding cost (Section 5.2)
    \STATE $Cost_{d_{l}} \leftarrow f(\model_{d_{l}}, \mathcal{X}^{c}) \quad // f(\cdot): Equation (\ref{encoding})$
    \STATE /* Update $d_{l}$
    \IF{$Cost_{d_{l}} < \text{MinCost}$}
    \STATE $\text{MinCost} \leftarrow Cost_{d_{l}}$
    \STATE $d_{l} \leftarrow d_{l} + 1$
    \ELSE
    \STATE \textbf{return} $\model_{d_{l}}$
    \ENDIF
    \ENDWHILE
    \end{algorithmic}
\end{algorithm}
\end{figure}

\subsection{Optimization}
Summarizing the descriptions so far, we have shown the overall procedure of learning \model\ while grouping locations with similar user activity characteristics in Algorithm \ref{alg1}.
The basic idea of the algorithm is that we search \model\ to minimize the encoding cost in Eq. (\ref{encoding}) while increasing the number of area groups $d_{l}$.

\section{Experiments and Results}
\subsection{Experimental settings}

\begin{table}[t]
    \centering
    \small
    \caption{Datsets}
    \begin{tabular}{cll} \toprule
    ID & Dataset & Query \\ \midrule
    \multirow{2}{*}{US\#1} & \multirow{2}{*}{E-commerce} & Amazon/Apple/BestBuy/Costco/Craigslist/Ebay/ \\
    & & Homedepot/Kohls/Macys/Target/Walmart\\
    \multirow{2}{*}{US\#2} & \multirow{2}{*}{VoD} & AppleTV/ESPN/HBO/Hulu/Netflix/Sling/ \\
    & & Vudu/YouTube\\
    \multirow{2}{*}{US\#3} & \multirow{2}{*}{Sweets} & Cake/Candy/Chocolate/Cookie/Cupcake/ \\
    & & Gum/Icecream/Pie/Pudding\\
    \multirow{2}{*}{US\#4} & \multirow{2}{*}{Facilities} & Aquarium/Bookstore/Gym/Library/Museum/ \\
    & & Theater/Zoo\\
    \multirow{2}{*}{World\#1} & \multirow{2}{*}{Music} & Beyonce/KatyPerry/LadyGaga/Maroon5/ \\
    & & StevieWonder/TaylorSwift\\
    \multirow{2}{*}{World\#2} & \multirow{2}{*}{SNS} & Facebook/LINE/Slack/Snapchat/Twitter/ \\
    & & Viber/WhatsApp\\
    \multirow{2}{*}{World\#3} & \multirow{2}{*}{Apparel} & \multirow{2}{*}{Gap/H\&M/Primark/Uniqlo/Zara} \\
    & & \\ \bottomrule
    \end{tabular}
    \label{query}
\end{table}

\subsubsection{Datasets}
We used the two kinds of online user activity data from different target areas on GoogleTrends, which contained weekly web search volumes collected for about 10 years, from January 1, 2011, to December 31, 2020.
\begin{itemize}
    \item \textbf{US data} includes web search volumes for 50 states of the US. Four datasets of this type were constructed: 
    \item \textbf{World data} includes three kinds of web search volumes for the top 50 countries ranked by GDP score. 
\end{itemize}
The queries of our datasets are described in Table~\ref{query}.

\subsubsection{Comparative models}
We used the following comparative models, which are state-of-the-art algorithms for modeling and forecasting time series:
\begin{itemize}
    \item \textbf{EcoWeb}~\cite{matsubara2015web}, which is intended for modeling online user activity data as well as our model, is a mathematical model constructed on the basis of differential equations.
    \item \textbf{SMF}~\cite{hooi2019smf} is a matrix factorization model that takes into account seasonal patterns.
    It is mainly used for time series forecasting and anomaly detection.
    \item \textbf{DeepAR}~\cite{salinas2020deepar} has an encoder-decoder structure that employs an auto-regression RNN modeling probabilistic distribution in the future. 
    Here, we selected a two-stack LSTM as the structure and its suitable number of hidden units in the validation period.
    \item \textbf{Gated Recurrent Unit (GRU)}~\cite{che2018recurrent} is a RNN-based model for time series forecasting.
    We employed an encoder-decoder architecture based on the GRU for the multi-step-ahead forecast. 
    We also applied a dropout rate of 0.5 to the connection of the output layer.
    \item \textbf{Informer}~\cite{zhou2021informer}, which is a transformer-based model based on ProbSparse self-attention and self-attention distilling, is known for its remarkable efficiency in long time series forecasting. We select a two-layer stack for both encoder and decoder and set 26 (half of the prediction sequence length) as the token length of the decoder.
\end{itemize}

\subsubsection{Setting}
We forecast the online user activity data using \model\ and other comparative models for the validation of the prediction performance.

\noindent \textbf{Dataset preprocessing}: 
We set search volumes in the datasets from January 1, 2011, to December 31, 2017, as the training and modeling period (seven years $ = t_{c}$), then from January 1, 2018, to December 31, 2018, as the validation period (one year), and then from January 1, 2019, to December 31, 2020, as the test period (two years).
We then normalized their search volumes in the range 0 to 1.

\noindent \textbf{Evaluation metrics}: 
We evaluate the predictive performance of these models at three points, 13 weeks (3 months), 26 weeks (half a year), and 52 weeks (a year) ahead.
Two evaluation metrics were utilized to compare the forecast performance levels of each model: root mean squared error ($\rm{RMSE}$) and mean absolute error ($\rm{MAE}$).
A lower value in these metrics indicates better forecasting accuracy.

\noindent \textbf{Hyper-parameter}: 
All the parameters in neural network based models were updated in the Adam update rule~\cite{kingma2014adam}.
We set $52$ as the input and forecast length and MSE loss as the loss function for DeepAR, GRU, and Informer.
We then conducted a grid search for parts of parameters such as hidden layers in comparative models.
In the training of \model, we selected the sizes of the RNN hidden layers in Eq. (\ref{react1}) as (16, 32, 64) in the validation period.
We also set $p$ in Eq. (\ref{season}) as $52$ to capture annual patterns from weekly datasets, the hyperparameters $\alpha$ and $\beta$ in Eq. (\ref{loss}) as $0.1$, and the number of epochs as 2,000 with early stopping.

\begin{table}[!tb]
  \centering
    \caption{Forecasting performance comparison}
    \small
\scalebox{0.9}{
\begin{tabular}{c|l|rrrrrr} \toprule
     & & \multicolumn{2}{c}{13 weeks} & \multicolumn{2}{c}{26 weeks} & \multicolumn{2}{c}{52 weeks}\\ \cmidrule(lr){3-4} \cmidrule(lr){5-6} \cmidrule(lr){7-8}
     Dataset & \multicolumn{1}{c|}{Model}  & $\rm{RMSE}$  & $\rm{MAE}$ & $\rm{RMSE}$ & $\rm{MAE}$ & $\rm{RMSE}$ & $\rm{MAE}$\\ \midrule

     \multirow{6}{*}{US\#1}& EcoWeb & 0.1470 & 0.0950 & 0.1554 & 0.1082 & 0.1654 & 0.1197\\
     & SMF & 0.0869 & 0.0620 & 0.0910 & 0.0654 & 0.1012 & 0.0674\\
     & DeepAR & 0.1003 & 0.0634 & 0.1302 & 0.0907 & 0.1385 & 0.1014\\
     & GRU &  0.1723 & 0.1175 & 0.1924 & 0.1374 & 0.2059 & 0.1525\\
     & Informer & 0.1477 & 0.1045 & 0.1375 & 0.0985 & 0.1575 & 0.1111\\
     & \model & \textbf{0.0478} & \textbf{0.0257} & \textbf{0.0574} & \textbf{0.0323} & \textbf{0.0631} & \textbf{0.0365}\\ \midrule

     \multirow{6}{*}{US\#2}& EcoWeb & 0.1440 & 0.1133 & 0.1981 & 0.1621 & 0.1920 & 0.1684\\
     & SMF & 0.0621 & 0.0445 & 0.0713 & 0.0522 & 0.0760 & 0.0529\\
     & DeepAR & 0.1471 & 0.1026 & 0.1781 & 0.1314 & 0.1906 & 0.1474\\
     & GRU &  0.1518 & 0.1171 & 0.1619 & 0.1231 & 0.1683 & 0.1440\\
     & Informer & 0.1277 & 0.0878 & 0.1292 & 0.0876 & 0.1436 & 0.1012\\
     & \model & \textbf{0.0245} & \textbf{0.0130} & \textbf{0.0276} & \textbf{0.0156} & \textbf{0.0310} & \textbf{0.0181}\\ \midrule

     \multirow{6}{*}{US\#3}& EcoWeb & 0.1555 & 0.1208 & 0.1730 & 0.1384 & 0.1754 & 0.1369\\
     & SMF & 0.0276 & 0.0186 & 0.0281 & 0.0170 & 0.0281 & 0.0190\\
     & DeepAR & 0.1107 & 0.0753 & 0.1267 & 0.0833 & 0.1309 & 0.0908\\
     & GRU &  0.1300 & 0.0869 & 0.1368 & 0.9843 & 0.1368 & 0.0939\\
     & Informer & 0.1322 & 0.0954 & 0.1311 & 0.0946 & 0.1279 & 0.0914\\
     & \model & \textbf{0.0200} & \textbf{0.0121} & \textbf{0.0222} & \textbf{0.0136} & \textbf{0.0238} & \textbf{0.0148}\\ \midrule

     \multirow{5}{*}{US\#4} & EcoWeb & 0.0847 & 0.0573 & 0.1348 & 0.0950 & 0.1511 & 0.1182\\
     & SMF & 0.0905 & 0.0762 & 0.1100 & 0.0751 & 0.1206 & 0.1077\\
     & DeepAR & 0.0927 & 0.0682 & 0.1662 & 0.1119 & 0.2168 & 0.1639\\
     & GRU &  0.1199 & 0.0872 & 0.1737 & 0.1223 & 0.2319 & 0.1764\\
     & Informer & 0.1014 & 0.0720 & 0.0992 & 0.0690 & 0.1055 & 0.0762\\
     & \model & \textbf{0.0495} & \textbf{0.0256} & \textbf{0.0610} & \textbf{0.0358} & \textbf{0.0794} & \textbf{0.0503}\\ \midrule

     \multirow{6}{*}{World\#1} & EcoWeb & 0.1259 & 0.0831 & 0.1422 & 0.1034 & 0.2101 & 0.1460\\
     & SMF & 0.0936 & 0.0783 & 0.0901 & 0.0602 & 0.1087 & 0.0787\\
     & DeepAR & 0.0900 & 0.0636 & 0.0929 & 0.0681 & 0.1395 & 0.0972\\
     & GRU & 0.0633 & 0.0452 & 0.0718 & 0.0501 & 0.0823 & 0.0572\\
     & Informer & 0.0704 & 0.0423 & 0.0719 & 0.0416 & 0.0738 & 0.0446\\
     & \model & \textbf{0.0454} & \textbf{0.0274} & \textbf{0.0477} & \textbf{0.0286} & \textbf{0.0546} & \textbf{0.0331}\\ \midrule

     \multirow{6}{*}{World\#2} & EcoWeb & 0.0908 & 0.0300 & 0.1089 & 0.0570 & 0.1353 & 0.0742\\
     & SMF & 0.0841 & 0.0436 & 0.0799 & 0.0454 & 0.0826 & 0.0480\\
     & DeepAR & 0.0374 & \textbf{0.0098} & \textbf{0.0585} & 0.0199 & 0.0643 & 0.0209\\
     & GRU &  0.0401 & 0.0159 & 0.0588 & \textbf{0.0174} & 0.0739 & 0.0254\\
     & Informer & \textbf{0.0371} & 0.0159 & 0.0595 & 0.0196 & \textbf{0.0642} & \textbf{0.0208}\\
     & \model & 0.0704 & 0.0271 & 0.0711 & 0.0304 & 0.0831 & 0.0351\\ \midrule

     \multirow{5}{*}{World\#3}& EcoWeb & 0.0523 & 0.0208 & 0.0626 & 0.0200 & 0.1080 & 0.0293\\
     & SMF & 0.0206 & 0.0111 & 0.0289 & 0.0160 & 0.0254 & 0.0195\\
     & DeepAR & 0.0211 & 0.0110 & 0.0275 & 0.0099 & 0.0613 & 0.0214\\
     & GRU &  0.0191 & 0.0090 & 0.0217 & \textbf{0.0096} & 0.00235 & 0.0115\\
     & Informer & 0.0223 & 0.0105 & 0.0226 & 0.0105 & \textbf{0.0214} & 0.0108 \\
     & \model & \textbf{0.0176} & \textbf{0.0085} & \textbf{0.0214} & \textbf{0.0096} & 0.0221 & \textbf{0.0100}\\ \bottomrule
  \end{tabular}
  }
    \label{reuslt}
\end{table}

\subsection{Results}
The experimental results are presented in Table~\ref{reuslt}.
These results indicate that \model\ outperformed most of the comparative models, confirming the benefits of our model architecture and algorithm.
EcoWeb, which is a linear model that captures only interactions between items to be built with the same motivation as \model, was unstable for long-ahead forecasting.
SMF, which focused on the extraction of seasonal patterns, outputs a good result of forecasting online user activity data with strong seasonality, compared to EcoWeb.
RNN-based models such as DeepAR and GRU were superior baseline models and exhibited almost the same performance as Informer in the 13 week forecast, but in some cases, they showed unstable results in long-ahead forecasting such as 52 weeks ahead.
These results indicate that it is not easy to improve the accuracy of long-ahead forecasts.
On the other hand, Informer, which is built for long-term forecasting, achieved the highest accuracy among the comparative models.
Informer showed a trend whereby the accuracy with the 52 weeks ahead forecast is not much worse than that with the 13 weeks ahead forecast.

\model\ achieved the highest accuracy in 5 out of 7 datasets for RMSE and 6 out of 7 for MAE compared to other models.
Our model shows high predictive performance as well as high interpretability by explicitly modeling the online user activity data with capturing the three hidden components: keyword interaction, location diffusion, and seasonality.

\begin{figure}[t!]
    \centering
    \includegraphics[width=\linewidth]{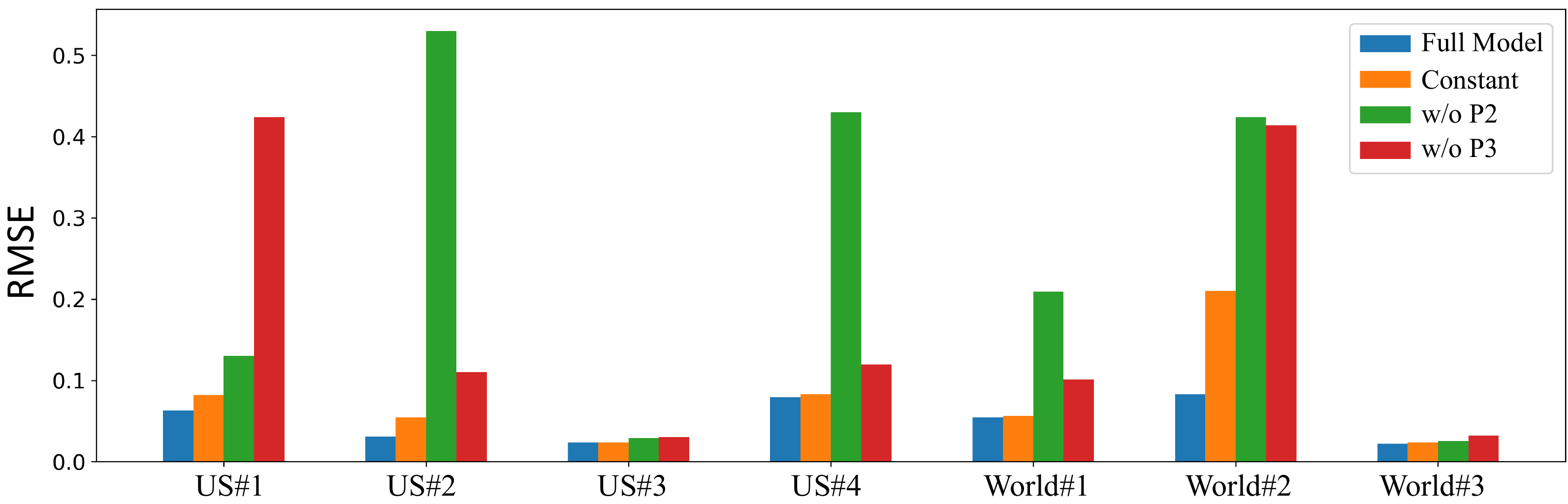}
    \caption{Experimental ablation results for RMSE performance of 52 weeks ahead forecasting.}
    \label{ablation}
\end{figure}

\noindent \textbf{Ablation study}: 
To demonstrate the effectiveness of all the components of \model, we performed ablation experiments with three ablation models, namely, Constant model, and Full model w/o P2, and w/o P3.
\textbf{Constant model} replaces the RNN in Eq. (\ref{react1}) in \model\ with time-invariant parameters. 
This reveals the benefit of time-varying components.
\textbf{Full model w/o P2} and \textbf{Full model w/o P3} are the removals of the diffusion term and the seasonal term in \model\, respectively.
Figure~\ref{ablation} shows the RMSE performances of 52 weeks ahead forecasting in each dataset.
Full model outperformed the other ablation models in all datasets.
Full model w/o P2 and w/o P3, which exclude the diffusion term and seasonal term, respectively, exhibited significantly poorer forecasting performance.
Also, Constant showed a reduced forecasting performance, suggesting the effectiveness of the time-varying parameters generated by RNN.
Each component in \model\ is useful for modeling online user activity data.

\begin{figure*}[t]
  
  \begin{subsubcaption}
    \begin{subfigure}[b]{0.35\linewidth}
    \centering
    \includegraphics[width=\textwidth]{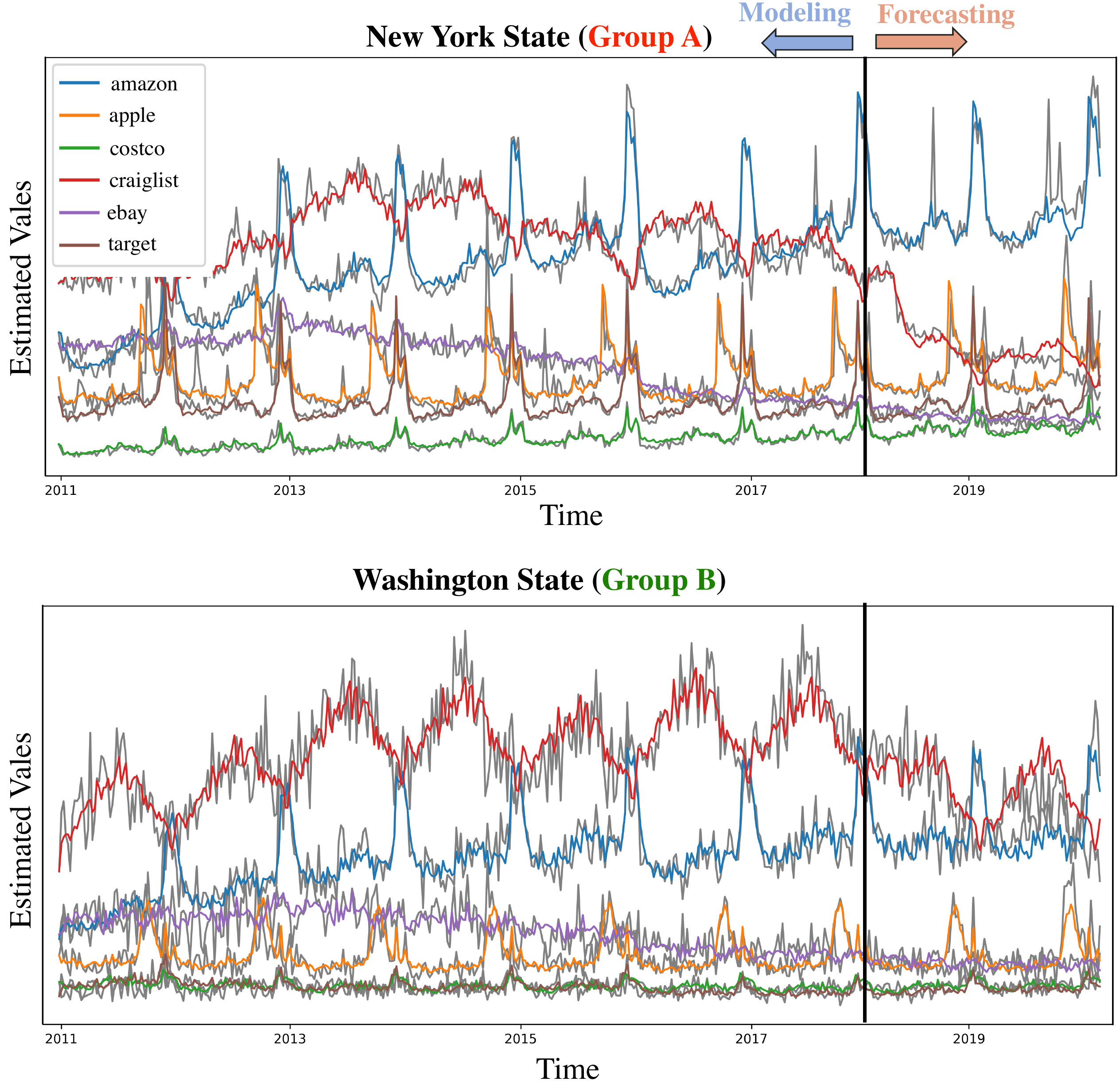}
    \subcaption{Modeling and forecasting results}
  \end{subfigure}
  \hfill
  \begin{subfigure}[b]{0.16\linewidth}
    \centering
    \includegraphics[width=\textwidth]{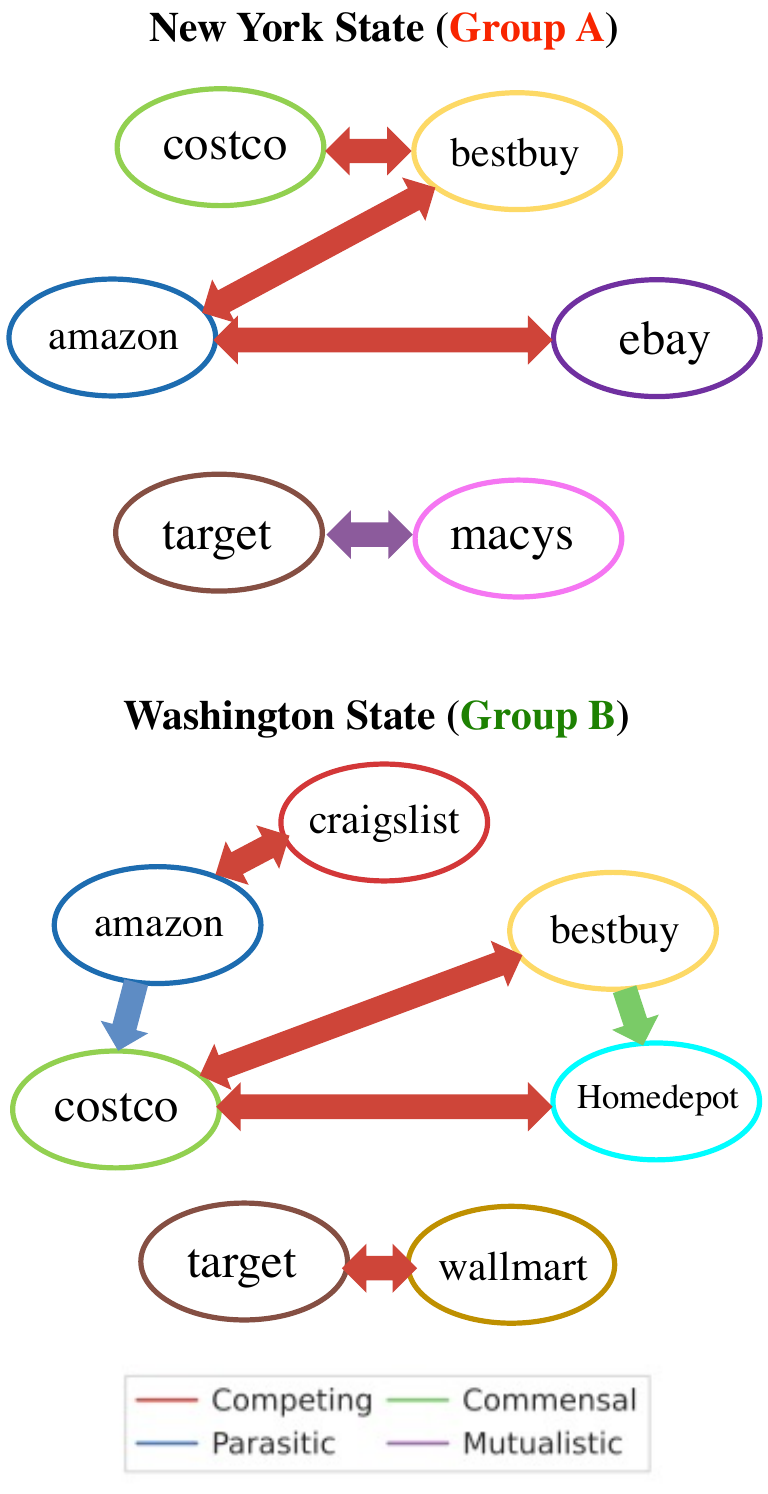}
    \subcaption{Interactions}
  \end{subfigure}
  \hfill
  \begin{subfigure}[b]{0.46\linewidth}
    \centering
    \includegraphics[width=\textwidth]{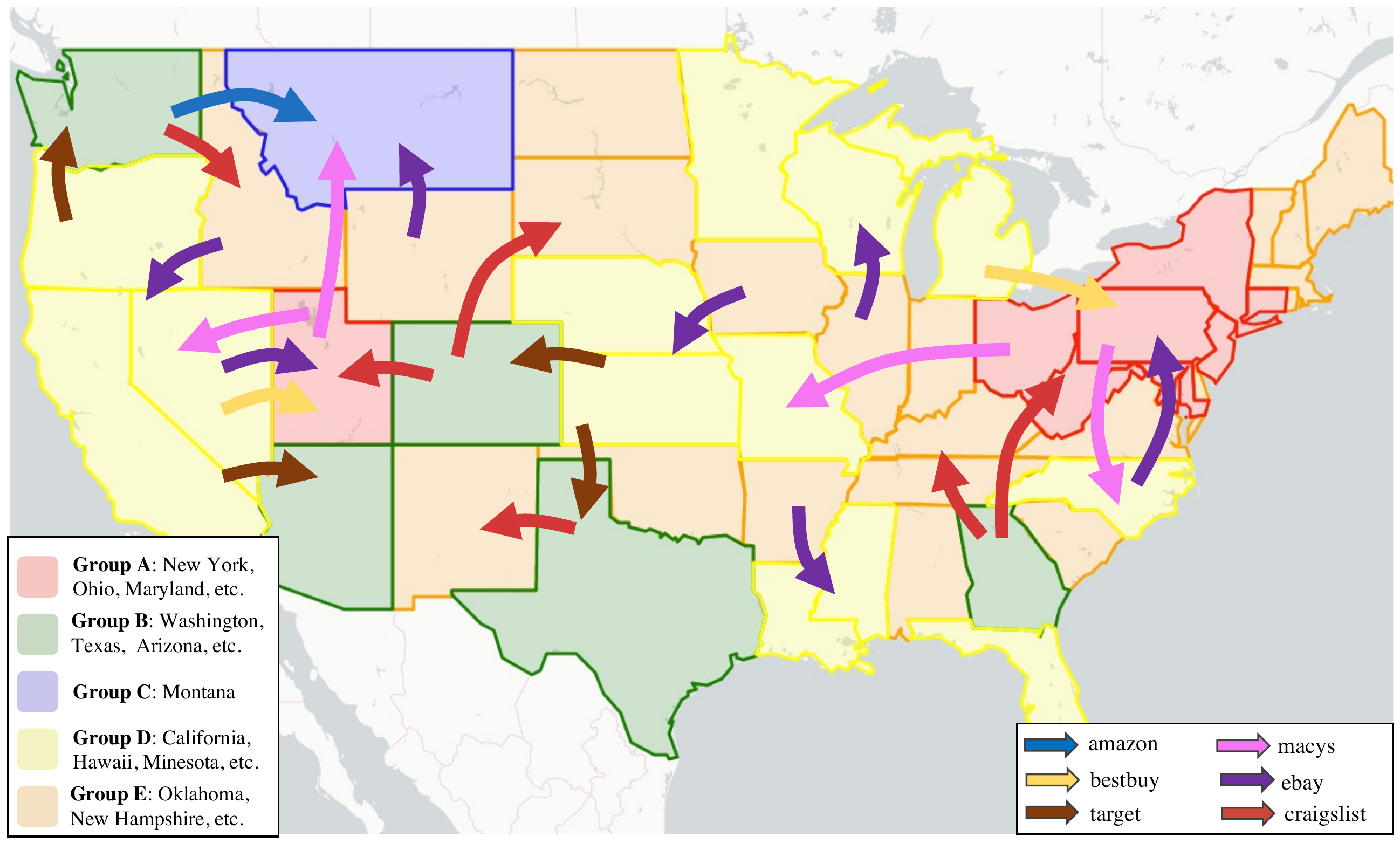}
    \subcaption{Diffusion process of each area group}
  \end{subfigure}
  \end{subsubcaption}
  
 \vspace{10pt}
 \begin{subsubcaption}
  \begin{subfigure}[b]{0.35\linewidth}
    \centering
    \includegraphics[width=\textwidth]{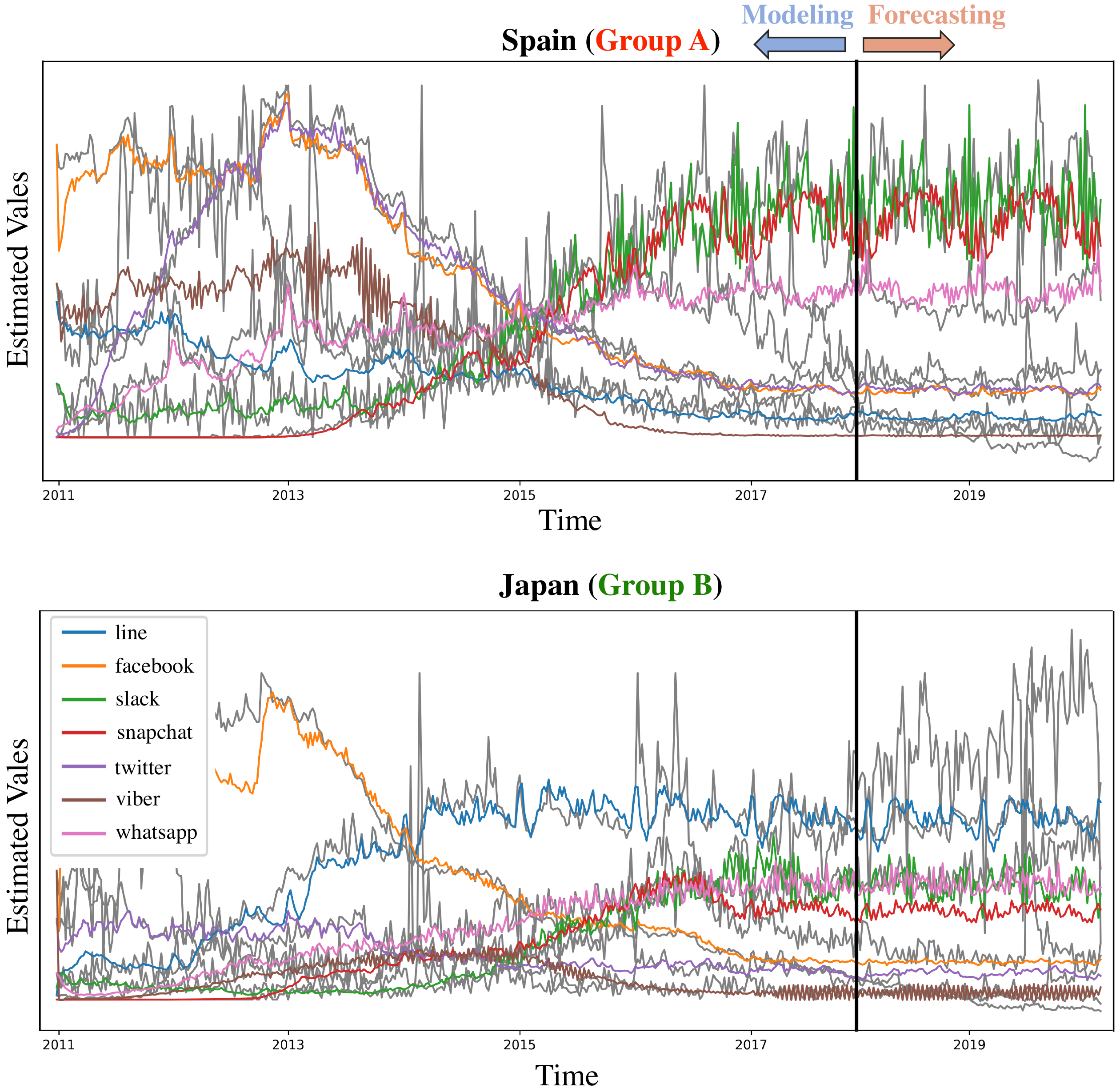}
    \subcaption{Modeling and forecasting results}
  \end{subfigure}
  \hfill
  \begin{subfigure}[b]{0.16\linewidth}
    \centering
    \includegraphics[width=\textwidth]{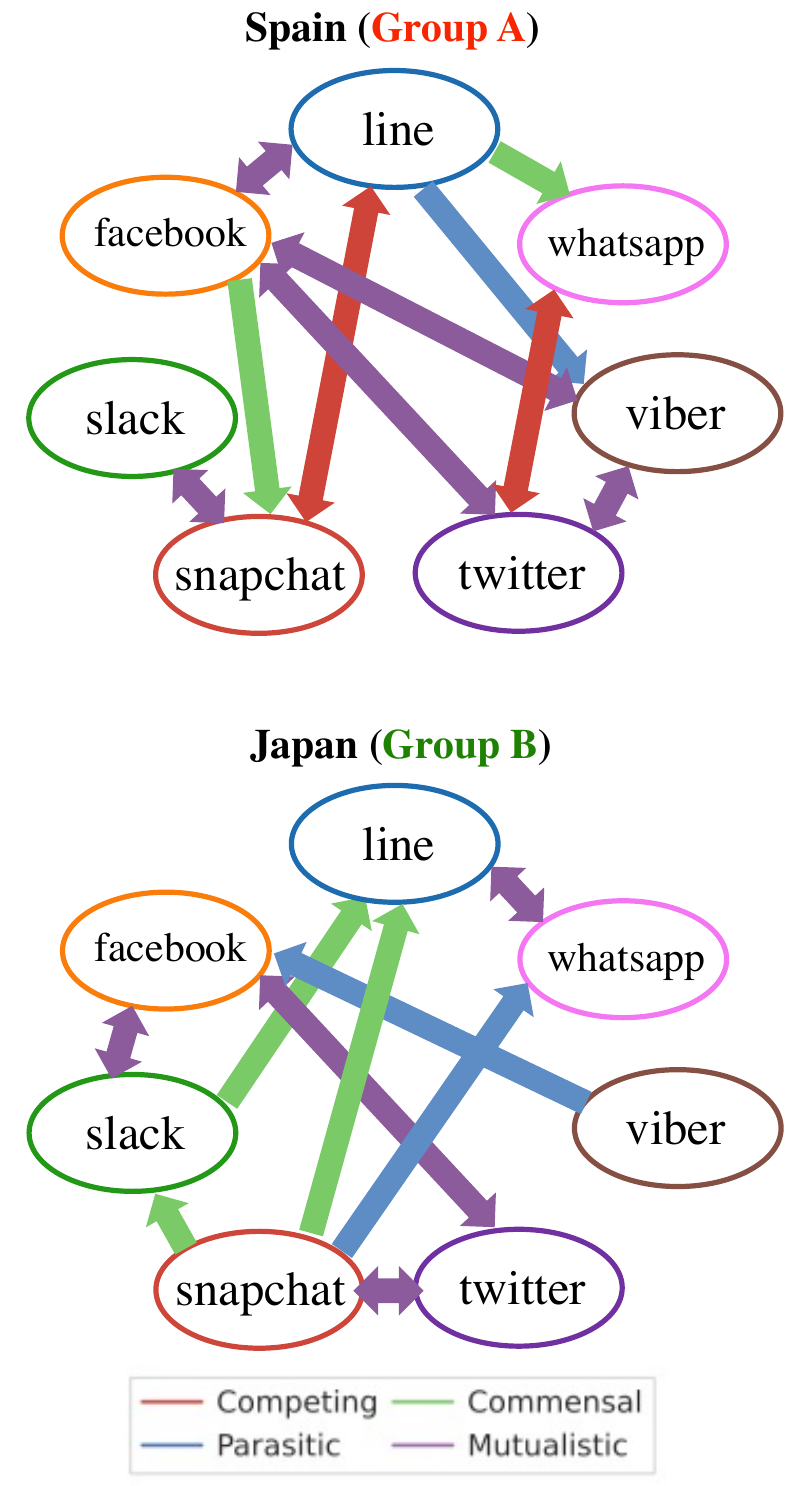}
    \subcaption{Interactions}
  \end{subfigure}
  \hfill
  \begin{subfigure}[b]{0.46\linewidth}
    \centering
    \includegraphics[width=\textwidth]{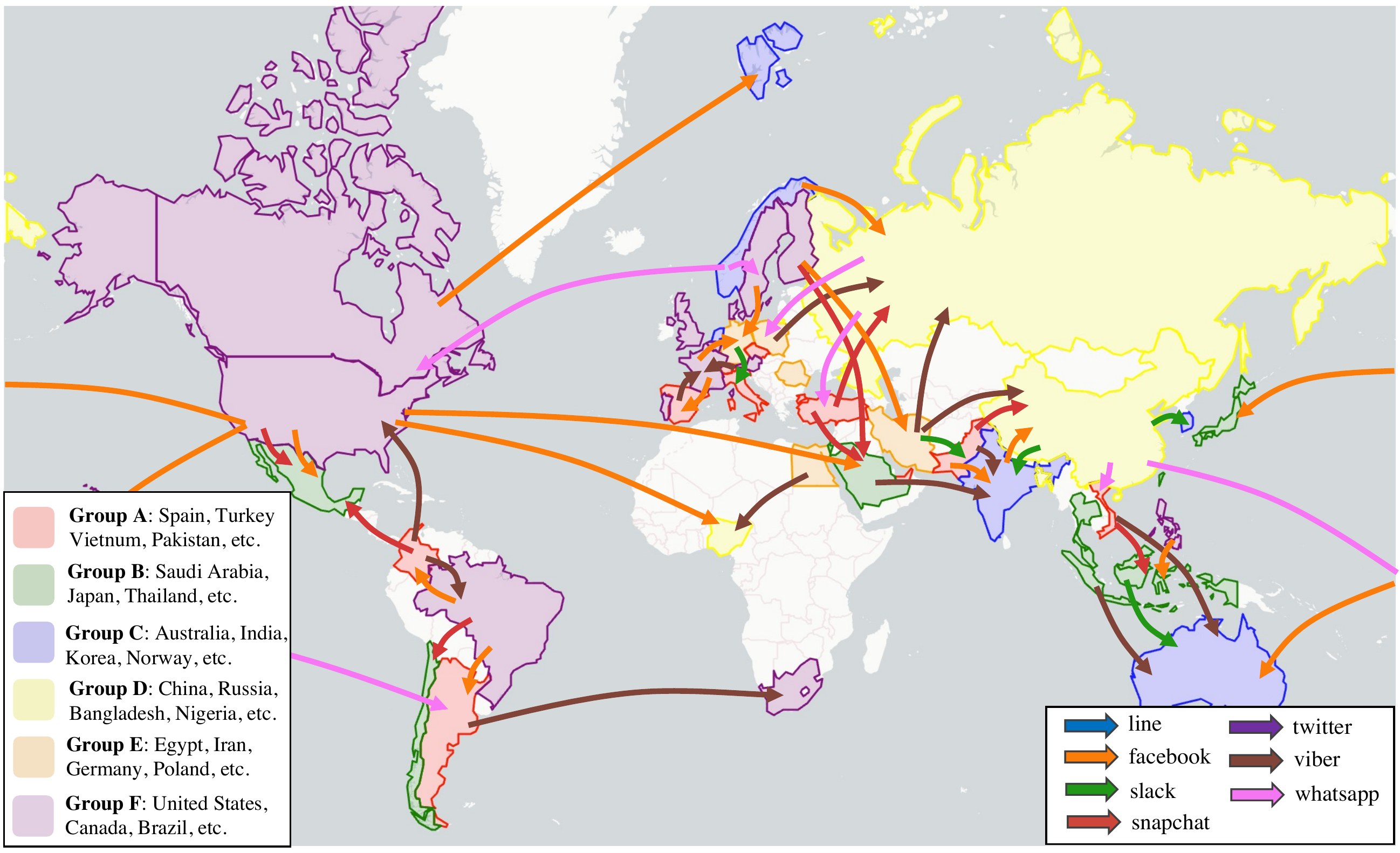}
    \subcaption{Diffusion process of each area group}
  \end{subfigure}
  \end{subsubcaption}
  \caption{\model\ modeling and forecasting results for US\#1 (a) and World\#2 (b) data: (a/b-1) shows the modeling and forecasting results of \model\ given the observational data. It represents observational data in gray and the modeling and forecasting results for each keyword in various colors. 
  (a/b-2) shows the network graphs summarizing four latent interactions between keywords in each country. 
  \model\ uncovers four types of interaction, each of which is represented by arrows. 
  Competing and mutualistic relationships are bidirectional, while commensal and parasitic relationships are one-directional: source species benefit from destination species’ growth. 
  (a/b-3) shows the clustering results with countries/states divided into multiple area groups and the flow of influences between each area group.
  }
  \label{casesutdy}
\end{figure*}

\section{Case study}
Here, we describe the analytical results obtained with \model.
The result of World\#1 has already been presented in Section 1 (i.e., Figure~\ref{overview_figure} and \ref{overview_figure2}.)
Figure~\ref{casesutdy} (a) and (b) show the \model\ modeling and forecasting results for US\#1 and World\#2 tensor data, respectively.
The figure contains three results obtained by \model\ modeling each set of data: the modeling and forecasting results of the tensor data, the latent interactions between keywords for each country obtained from $\boldsymbol{C}$, and the location clustering results and the flow of influences between each area group obtained from $\boldsymbol{D}$.

\noindent \textbf{Case study of US\#1}: 
Figure~\ref{casesutdy} (a-1) shows that our proposed model successfully captures and forecasts latent dynamic patterns for multiple countries and keywords.
Concretely, our proposed model can capture the seasonality patterns in amazon and apple, as well as the decreasing trends of craigslist.
Figure~\ref{casesutdy} (a-2) based on the reaction term in \model\ indicates the network graphs summarizing latent interactions between keywords with different colors.
For example, in New York State, it shows that amazon competes with bestbuy and ebay, and target and macys are a mutualistic relationship.
Figure~\ref{casesutdy} (a-3) shows the results of clustering each state into multiple groups.
The influence of macys from group A, into which \model\ groups the northeast region of the US, affects other areas.

\noindent \textbf{Case study of World\#2}:
Figure~\ref{casesutdy} (b-1) shows that \model\ captures the increasing trend of slack and snapchat in Spain and the decreasing trend of facebook in Japan.
On the other hand, it failed to adequately capture the trend of slack, which increased rapidly in popularity in Japan after 2018.
As shown in Figure~\ref{casesutdy} (b-2), \model\ discovers many commensal and mutualistic relationships between social media platforms: e.g., in Spain, facebook has mutualistic relationships with several platforms such as line, viber, and twitter.
We find that many social media tend to be symbiotic rather than compete for online users.
Figure~\ref{casesutdy} (b-3) shows that our model groups countries such as Japan and Saudi Arabia, where twitter is mainly used, into group B and countries such as the United States and the United Kingdom, where facebook is mainly used, into group F.
Our results about the influence interactions between area groups show that the influence of facebook flows from group F to other area groups.
Also, the influence of viber founded in Israel flows from group E, including countries in the Middle East, to other area groups.

As our case studies show, \model\ provides the ability to describe complex interactions that can reveal underlying relationships among keywords and locations, and is suitable for modeling and forecasting online user activity data.

\section{Conclusion}
In this study, we attempted to model and forecast large time-evolving online user activity data such as web search volumes.
To achieve this, we proposed an effective modeling and forecasting method, namely \model\, based on reaction-diffusion and ecological systems.
Our method can recognize trends, seasonality and interactions in input observations by extracting their latent dynamic systems.
Throughout the search volume forecasting experiments in multiple GoogleTrends datasets, the proposed model achieved the highest performance by capturing the latent dynamics.

In addition to its predictive performance, \model\ provides the latent interactions and the influence flows hidden behind observational data in a human-interpretable form.
For example, in Washington State, costco competes with bestbuy and homedepot for online users, as shown by our case studies.
Uncovering such latent interactions behind the observational data assists to our decision-making.

\begin{acks}
The authors would like to thank the anonymous referees for their valuable comments and helpful suggestions. 
This work was supported by JSPS KAKENHI Grant-in-Aid for Scientific Research Number JP20H00585, JP21H03446, NICT 21481014, MIC/SCOPE 192107004, JST-AIP JPMJCR21U4, and ERCA-Environment Research and Technology Development Fund JPMEERF20201R02.
\end{acks}

\bibliographystyle{ACM-Reference-Format}
\bibliography{sample-base}

\end{document}